%% file: main.tex
\documentclass[manuscript,screen]{acmart}

\usepackage{array}
\usepackage%[disable]	% uncomment to hide all the notes
	{todonotes} 

\usepackage{textcomp}
\usepackage{stfloats}
\usepackage{verbatim}

\usepackage{svg}

\usepackage[utf8]{inputenc}
\usepackage{algorithm}
\usepackage[noend]{algpseudocode}

\usepackage{natbib}
\usepackage{tabto}
\usepackage{graphicx}
\usepackage{textcomp}
\usepackage{xcolor}
\usepackage[normalem]{ulem}
\usepackage[per-mode=symbol]{siunitx}
\def\defaultDigitGrouping{5}% To make digit grouping only after 4 digits, not exactly with 4.
\usepackage{mathtools}

\usepackage{booktabs}% To use extra commands to format tables (\toprule, \midrule and \bottomrule).
\usepackage{soul}
\usepackage{multirow}
\usepackage{url}
\usepackage{glossaries}
\usepackage{mdwlist}		% Defines itemize*, enumerate* y description*, which reduce vertical spacing.
\usepackage{placeins}
\usepackage{tablefootnote}
\hypersetup{colorlinks=false,pdfborder={0 0 0},hypertexnames=false}
\usepackage{float}
\usepackage{subcaption}

\usepackage{svg}
\usepackage{amsmath}

\newacronym{api}{API}{application programming interface}
\newacronym{dsp}{DSP}{digital signal processing}
\newacronym{dvfs}{DVFS}{dynamic voltage frequency scaling}
\newacronym{hal}{HAL}{hardware abstraction layer}
\newacronym{mac}{MAC}{multiply-accumulate}
\newacronym{ml}{ML}{machine learning}
\newacronym{muapt}{MUAPT}{motor unit action potential train}
\newacronym{sdk}{SDK}{software development kit}
\newacronym{ulp}{ULP}{ultra-low power}
\newacronym{nda}{NDA}{non-disclosure agreement}
\newacronym{os}{OS}{operating system}
\newacronym{fxp}{FxP}{fixed-point}
\newacronym{fp}{FP}{floating-point}
\newacronym{soa}{SoA}{state-of-the-art}
\newacronym{isa}{ISA}{Instruction Set Architecture}

\newacronym{eembc}{EEMBC}{Embedded Microprocessor Benchmark Consortium}
\newacronym{iot}{IoT}{Internet-of-Things}
\newacronym{iomt}{IoMT}{Internet-of-Medical-Things}
\newacronym{wcet}{WCET}{worst-case execution time}

\newacronym{seizdetcnn}{SeizureDetCNN}{seizure detector convolutional neural network}
\newacronym{seizdetsvm}{SeizureDetSVM}{seizure detector support vector machine}
\newacronym{coughdet}{CoughDet}{cough detector}
\newacronym{cwm}{CognWorkMon}{cognitive workload monitor}
\newacronym{gcl}{GestureClass}{gesture classifier}
\newacronym{hcl}{HeartBeatClass}{heartbeat classifier}
\newacronym{ecl}{EmotionClass}{emotion classifier}
\newacronym{bpfree}{Bio-BPfree}{biological back-propagation-free}
\newacronym{semg}{sEMG}{surface electromyography}
\newacronym{ecg}{ECG}{electrocardiogram}
\newacronym{eeg}{EEG}{electroencephalogram}
\newacronym{gsr}{GSR}{galvanic skin response}
\newacronym{ppg}{PPG}{photoplethysmography}
\newacronym{st}{ST}{skin temperature}

\newacronym{bss}{BSS}{blind source separation}
\newacronym{cnn}{CNN}{convolutional neural network}
\newacronym{rp}{RP}{random projections}
\newacronym{rf}{RF}{random forest}
\newacronym{mf}{MF}{morphological filter}
\newacronym{ma}{MAVG}{moving average}
\newacronym{bpf}{BPF}{band-pass filter}
\newacronym{lpf}{LPF}{low-pass filter}
\newacronym{rms}{RMS}{root-mean-square}
\newacronym{fft}{FFT}{fast Fourier transform}
\newacronym{relen}{Rel-En}{realtive energy}
\newacronym{dnn}{DNN}{deep neural network}
\newacronym{svm}{SVM}{support vector machine}
\newacronym{ica}{ICA}{independent component analysis}
\newacronym{mlp}{MLP}{multilayer Perceptron}
\newacronym{knn}{KNN}{k-nearest neighbors}
\newacronym{mfcc}{MFCC}{mel-frequency cepstral coefficients}
\newacronym{ecgfpdel}{ECG-FPDEL}{ECG fiducial points delineation}
\newacronym{avg}{AVG}{average}
\newacronym{reward}{REWARD}{Relative-Energy-based WeArable R-peak Detection}
\newacronym{rri}{RRI}{R-peak interval}
\newacronym{edr}{EDR}{ECG-derived respiration}
\newacronym{hrv}{HRV}{heart-rate variability}
\newacronym{lpc}{LPC}{linear predictive coefficients}
\newacronym{plomb}{PLOMB}{Lomb-Scargle periodogram}
\newacronym{adam}{ADAM}{Adam optimizer}
\newacronym{convgrad}{CONV-GRAD}{Convolutional block gradients}
\newacronym{dct}{DCT}{Discrete Cosine Transform}
\newacronym{pt}{PT}{Pan-Tompkin}
\newacronym{lut}{LUT}{look-up table}

\newacronym{dma}{DMA}{direct memory access}
\newacronym{fc}{FC}{fabric controller}
\newacronym{fpu}{FPU}{floating-point unit}
\newacronym{mcu}{MCU}{micro-controller unit}
\newacronym{pmu}{PMU}{power management unit}
\newacronym{spi}{SPI}{serial peripheral interface}
\newacronym{adc}{ADC}{analog-to-digital converter}
\newacronym{imu}{IMU}{inertial measurement unit}
\newacronym{ic}{IC}{integrated circuit}
\newacronym{ldo}{LDO}{low-dropout regulator}

\usepackage{amsmath,amsfonts}

\usepackage{array}
\usepackage{multirow}

\renewcommand{\eqref}[1]{(\ref{#1})}

\newcommand{\figref}[1]{\mbox{Figure~\ref{#1}}}
\newcommand{\tblref}[1]{\mbox{Table~\ref{#1}}}

\AtBeginDocument{%
  \providecommand\BibTeX{{%
    \normalfont B\kern-0.5em{\scshape i\kern-0.25em b}\kern-0.8em\TeX}}}

\setcopyright{acmcopyright}
\copyrightyear{2024}
\acmYear{2024}
\acmDOI{XXXXXXX.XXXXXXX}

\begin{document}

\title{Enabling Efficient Wearables: An Analysis of Low-Power Microcontrollers for Biomedical Applications}

% \author{Author list omitted for blind review purposes}

\author{Dimitrios~Samakovlis}
\affiliation{%
    \institution{Embedded Systems Laboratory, École Polytechnique Fédérale de Lausanne (EPFL)}
    \city{Lausanne, VD}
    \country{Switzerland}
}
\email{dimitrios.samakovlis@epfl.ch}

\author{Stefano~Albini}
\affiliation{%
    \institution{Embedded Systems Laboratory, École Polytechnique Fédérale de Lausanne (EPFL)}
    \city{Lausanne, VD}
    \country{Switzerland}
}

\author{Rubén~Rodríguez~Álvarez}
\affiliation{%
    \institution{Embedded Systems Laboratory, École Polytechnique Fédérale de Lausanne (EPFL)}
    \city{Lausanne, VD}
    \country{Switzerland}
}

\author{Denisa-Andreea~Constantinescu}
\affiliation{%
    \institution{Embedded Systems Laboratory, École Polytechnique Fédérale de Lausanne (EPFL)}
    \city{Lausanne, VD}
    \country{Switzerland}
}

\author{Pasquale~Davide~Schiavone}
\affiliation{%
    \institution{Embedded Systems Laboratory, École Polytechnique Fédérale de Lausanne (EPFL)}
    \city{Lausanne, VD}
    \country{Switzerland}
}

\author{Miguel~Peón-Quirós}
\affiliation{%
    \institution{EcoCloud Center, École Polytechnique Fédérale de Lausanne (EPFL)}
    \city{Lausanne, VD}
    \country{Switzerland}
}

\author{David~Atienza}
\affiliation{%
    \institution{Embedded Systems Laboratory, École Polytechnique Fédérale de Lausanne (EPFL)}
    \city{Lausanne, VD}
    \country{Switzerland}
}
\affiliation{%
    \institution{EcoCloud Center, École Polytechnique Fédérale de Lausanne (EPFL)}
    \city{Lausanne, VD}
    \country{Switzerland}
}

\renewcommand{\shortauthors}{Samakovlis, Dimitrios, et al.}

\begin{abstract}

Breakthroughs in ultra-low-power chip technology are transforming biomedical wearables, making it possible to monitor patients in real time with devices operating on mere \SI{}{\micro\watt}. Although many studies have examined the power performance of commercial microcontrollers, it remains unclear which ones perform best across diverse application profiles and which hardware features are most crucial for minimizing energy consumption under varying computational loads. Identifying these features for typical wearable applications and understanding their effects on performance and energy efficiency are essential for optimizing deployment strategies and informing future hardware designs. In this work, we conduct an in-depth study of \gls{soa} \glspl{mcu} in terms of processing capability and energy efficiency using representative end-to-end \gls{soa} wearable applications. We systematically benchmark each platform across three primary application phases: idle, data acquisition, and processing, allowing a holistic assessment of the platform processing capability and overall energy efficiency across varying patient-monitoring application profiles. Our detailed analysis of performance and energy discrepancies across different platforms reveals key strengths and limitations of the current low-power hardware design and pinpoints the strengths and weaknesses of \gls{soa} \glspl{mcu}. We conclude with actionable insights for wearable application designers and hardware engineers, aiming to inform future hardware design improvements and support optimal platform selection for energy-constrained biomedical applications.
\end{abstract}


\received{Received date}
\received[revised]{Revised date}
\received[accepted]{Accepted date}

%%
%% This command processes the author, the affiliation, and the title
\maketitle

\input{Introduction}
\input{RelatedWork}

\input{Applications}
\input{Experimental_setup}
\input{Experimental_results}
\input{Conclusions}
%\input{Acknowledgments.tex}%Removed for blind review
%\FloatBarrier

%%
%% The next two lines define the bibliography style to be used and
\bibliographystyle{ACM-Reference-Format}
\bibliography{bibliography}

\end{document}

%% file: Introduction.tex
\section{Introduction}\label{sec:intro}

Rapid advancement in wearable technology has revolutionized patient monitoring, offering unprecedented opportunities to improve healthcare delivery. Wearable systems promise to monitor chronic diseases through real-time patient monitoring. However, the wearable domain poses several challenges for application and hardware designers. The main challenge is increasing the computational complexity and prediction accuracy of algorithms while maintaining acceptable energy consumption and inference latency. 

Several hardware solutions have been proposed in the literature to facilitate the deployment of real-time biomedical algorithms. Typically, application developers opt for commercial low-power microcontrollers (e.g., STM32~\cite{stmulp}, GAP~\cite{gap_processors}) to deploy their applications. However, the selection of deployment platforms is usually based on weak criteria, as there is no comprehensive study on the processing and energy efficiency of different \glspl{mcu}. It is unclear which characteristics of the hardware design mainly impact performance and energy across different application profiles and workloads. Pinpointing critical hardware features by application type and understanding their impact on performance and energy is the key to achieving efficient application deployment and can be instructive for future hardware designs. 

An in-depth analysis of existing wearable hardware designs offers significant value in advancing future biomedical wearable designs. By thoroughly evaluating the performance, energy efficiency, and architectural choices of current low-power microcontroller platforms, researchers can gain crucial insights into the strengths and weaknesses of existing solutions. By critically examining shortcomings in power consumption and processing efficiency, hardware designers can develop targeted solutions to overcome these barriers. Equally important is recognizing and building on the strengths of existing \gls{mcu} architectures. The study of innovations in low power consumption can inspire researchers to refine and enhance them to create even more sophisticated wearable devices. This iterative learning and improvement process fosters a cycle of innovation, accelerating the development of smart wearable systems.

In this paper, we evaluate the processing and energy efficiency of five \gls{soa} commercial low-power platforms using end-to-end \gls{soa} wearable applications. We aim to increase awareness of critical applications and hardware design characteristics and their impact on performance and energy in the biomedical wearable domain. We evaluate all platforms during the three application phases: idle, acquisition, and processing. Our in-depth comparative analysis of the experimental results leads to important takeaways that provide insight and guide future hardware and application researchers. The contribution of this paper is threefold:
\begin{itemize}
  \item It offers guidance on wearable hardware design by identifying vital architectural features that influence processing performance and energy efficiency in \gls{soa} real-time biomedical applications.

  \item It provides guidelines to application designers on selecting the most suitable deployment platform for their application type based on the strengths and weaknesses of \gls{soa} platforms.

  \item It offers biomedical software developers insights into the application deployment process and the effects of specific biomedical workloads on wearable hardware performance.
  
\end{itemize}

 \begin{figure}[tp]
	\centering
	\includegraphics[width=\linewidth]{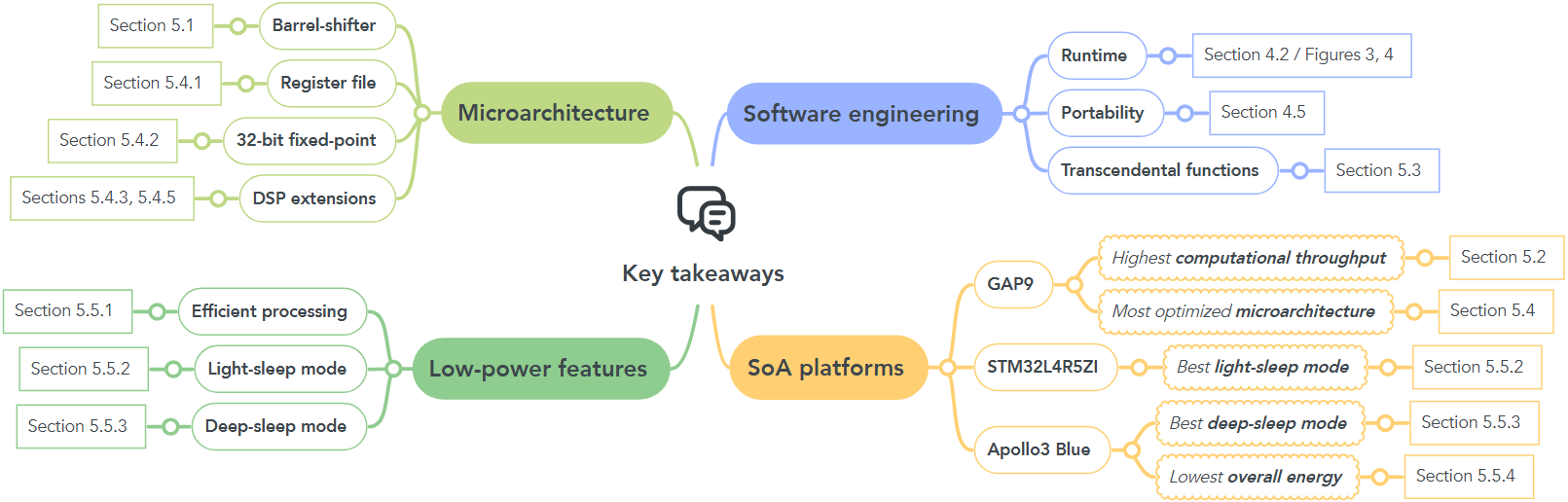}
	\caption{Map of the key takeaways per topic.}
	\label{fig:takeaways}
\end{figure}

Fig.~\ref{fig:takeaways} presents a map of the key takeaways for each topic covered in our study. The takeaways are the result of insight gained through the deployment process and the detailed analysis of different experiments. We have hierarchically categorized these takeaways according to the target area and have provided a link to the corresponding section that provided insight into the topic. 

We organize the remainder of this work as follows. In Section~\ref{sec:RelatedWork}, we compare works related to ours. In Section~\ref{sec:applications}, we showcase the benchmark applications considered for the evaluation of the platform. In Section~\ref{sec:ExpSetUp}, we describe the setup of our experiments on state-of-the-art low-power platforms. In Section~\ref{sec:ExpRes}, we analyze the results of running our experiments and summarize the key takeaways. Finally, we summarize the key findings of this work in Section~\ref{sec:conclusions}.

%% file: RelatedWork.tex
\section{Related Work}\label{sec:RelatedWork}

The authors of \cite{mcu_comparison_paper} summarize the architecture, peripherals, development environment, and application areas of five commercial \glspl{mcu}: AVR, 8052, PIC, ESP32, and STM32. The scope of that paper is to provide a high-level view of the characteristics of different board families and guide the embedded systems' community to a suitable platform selection for their needs. In contrast, our paper explicitly compares the \glspl{mcu}' critical hardware features that impact performance by directly comparing energy and processing capability while running real-time patient monitoring applications. Moreover, our considered platforms (i.e., STM32, GAP8, GAP9, Apollo3 Blue, and RP2040) are different.

The authors of BiomedBench~\cite{biomedbench} compare low-power \glspl{mcu} in terms of processing and energy efficiency under the biomedical workloads proposed in their benchmark suite. Their selected platforms match the platforms we considered for this study (i.e., STM32, GAP8, GAP9, Apollo3 Blue, and RP2040). Their superficial comparison reports performance and energy numbers, but does not delve into the underlying factors that lead to the reported results. In contrast, our study explores in depth the microarchitectural and architectural features that lead to performance discrepancies between platforms, the ability of platforms to tackle real-time constraints, and the impact of multi-core architectures. In addition, our study offers information on the deployment process and the effect of specific biomedical kernels on hardware performance.

To the best of our knowledge, no other scientific work has thoroughly explored the significant architectural features of low-power \glspl{mcu} under \gls{soa} biomedical applications. The insights gained through this in-depth comparison and deployment process can significantly benefit future hardware designers and application developers in the biomedical wearable domain, as shown in Section~\ref{sec:ExpRes:Summary}. Overall, this study not only offers an in-depth comparison, but also provides valuable guidelines for designing such platforms and developing software efficiently according to the platforms' characteristics.

%% file: Applications.tex
\section{Benchmark Applications}\label{sec:applications}

This section presents the benchmark applications considered for the evaluation of wearable platforms. We opt for the benchmark applications presented in BiomedBench~\cite{biomedbench}. BiomedBench is a \gls{soa} biomedical benchmark suite for wearables and offers representative computational workloads and application profiles for complete platform evaluation in the three phases (i.e., processing, acquisition, and idle).

Section~\ref{sec:applications:app_profile} describes the typical structure of wearable applications. In Section~\ref{sec:applications:benchmarks}, we summarize the selected benchmark applications, their main kernels, and their main characteristics.

 \begin{figure}[tp]
	\centering
	\includegraphics[width=.6\linewidth]{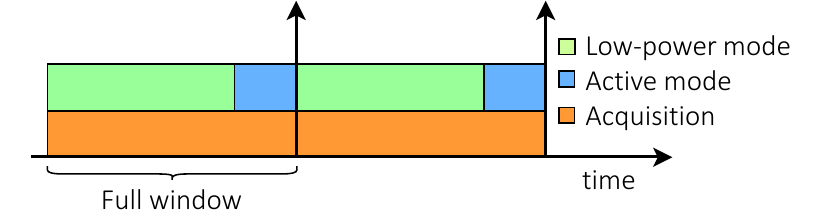}
	\caption{Device operating modes during two full application cycles.}
	\label{fig:appAcquisitionIdlePeriods}
\end{figure}

\subsection{Application profile}\label{sec:applications:app_profile}

 % Fig.~\ref{fig:appModularStructure} shows a 
 A typical biosignal monitoring application pipeline consists of several phases. First, sensors on the patient's body capture the biosignal, which is then forwarded as input to the processing device. The application processing phase takes place on the device. Despite the different pathologies and algorithms, the processing pipeline typically consists of three stages: signal preprocessing (i.e., filtering), feature extraction (i.e., time- or/and frequency-domain characteristics), and \gls{ml} inference. If the signal is classified as abnormal, the device triggers an alarm to notify the appropriate people and entities (i.e., through a smartphone).
 
Fig.~\ref{fig:appAcquisitionIdlePeriods} illustrates the operating modes of the processing device during two full application cycles of always-on acquisition. A full window consists of an idle phase during which the processing unit is in low-power mode (i.e., deep-sleep mode when not collecting samples and light-sleep mode when collecting samples) and a computation phase upon acquisition of a full window signal. The duration of the idle period can vary significantly between applications and can dominate system energy consumption.

\begin{table*}[tp]
\begin{center}
\resizebox{0.9\textwidth}{!}{%
\begin{tabular}{l@{\hskip0.25cm}l@{\hskip0.15cm}l@{\hskip0.15cm}r@{\hskip0.15cm}r@{\hskip0.15cm}r@{\hskip0.15cm}} 
    \toprule
    \textbf{Application} &  \textbf{Main operations} & \textbf{Duty Cycle} & \multicolumn{1}{l}{\textbf{Input bandwidth} } & \multicolumn{1}{l}{\textbf{Static data}} & \multicolumn{1}{r}{\textbf{Dynamic data}} \\
                &            &                &     \multicolumn{1}{c}{\textbf{\si{(\byte/\sec)}}}       & \multicolumn{1}{c}{\textbf{(\si{\kibi\byte})}} & \multicolumn{1}{c}{\textbf{(\si{\kibi\byte})}} \\
    \midrule
    \acrshort{hcl} & Branches (\acrshort{fxp} min/max search) & Low & 1536 & 25 & 30 \\ 
    \acrshort{seizdetsvm} & 32-bit \acrshort{fxp} multiplications/divisions & Very Low & 128 & 40 & 40 \\
    \acrshort{seizdetcnn} & 16-bit \acrshort{fxp} \acrshort{mac} & High & 11776 & 350 & 120 \\
    \acrshort{cwm} & 32-bit \acrshort{fxp} multiplications & Medium & 4096 & 90 & 50 \\
    \acrshort{gcl} &  32-bit \acrshort{fp} \acrshort{mac} & Very High & 192000 & 50 & 110 \\
    \acrshort{coughdet} & 32-bit \acrshort{fp} multiplications & Very High & 64400 & 568 & 160 \\
    \acrshort{ecl} & Branches (\acrshort{fp} sorting) & Low & 822 & 16 & 4 \\
    \acrshort{bpfree} & 32-bit \acrshort{fp} \acrshort{mac} & - & - & 1300 & 2600 \\
    \bottomrule
\end{tabular}
}
\end{center}
\caption{Benchmark applications' characterization~\cite{biomedbench}}
\label{App_Metrics_Table}
\end{table*}

\subsection{Benchmark applications}\label{sec:applications:benchmarks}
BiomedBench applications offer representative workloads and profiles for the processing, idle, and acquisition phases in the domain of low-power smart wearable systems. These applications feature different challenges during idle, acquisition, and processing, including varying active-to-idle ratios, input bandwidth, and computational kernels. This enables future users to thoroughly explore different parts of their architectures (e.g., sleep mode efficiency, digital signal processing efficiency, and signal acquisition efficiency).

\tblref{App_Metrics_Table} summarizes the main metrics of the applications, as presented in the original BiomedBench paper. The most important metrics for evaluating platforms' energy and processing efficiency are the type of main operations, the duty cycle, and the input bandwidth. The type of main operations defines the type of workload and hints at the microarchitectural design required to tackle it efficiently. The duty cycle defines the influence of the idle period on the overall energy footprint; a low duty cycle necessitates an efficient deep sleep mode. The input bandwidth quantifies the intensity of the signal acquisition; a high input bandwidth stresses the importance of an energy-efficient acquisition. In Section~\ref{sec:ExpRes}, we emphasize these observations through our in-depth analysis of \gls{soa} commercial \glspl{mcu} running these applications.

A brief description of the applications follows. For a more detailed explanation, refer to~\cite{biomedbench}.

\subsubsection{\Gls{hcl}}
The \gls{hcl}~\cite{heartbeatclassifier} uses the \gls{ecg} signal to classify heartbeats in 15-second windows. The dominant workload, the \gls{mf}, lies on the preprocessing step and takes more than \SI{80}{\percent} of the execution time. Its implementation uses a queue that constantly monitors the minimum and maximum elements of each sub-window. The application is coded in C with 16-bit fixed-point arithmetic and includes a multicore version.

\subsubsection{\Gls{seizdetsvm}}
The \gls{seizdetsvm}~\cite{Farnaz_SVM} works in \gls{ecg} to detect epileptic episodes in 60-second windows. The dominant kernel, the \gls{plomb}, lies on the feature extraction step and accounts for more than \SI{75}{\percent} of the execution time. The application is coded in C++ using 32-bit \gls{fxp} arithmetic. Hence, most operations are 32-bit \gls{fxp} multiplications with 64-bit intermediate results and 32-bit \gls{fxp} divisions with 64-bit nominators. The application also includes a multicore version.

\subsubsection{\Gls{seizdetcnn}}
The \gls{seizdetcnn}~\cite{Gmez2020AutomaticSD} works on \gls{eeg} to detect epileptic episodes in a 4-second window. There is no signal preprocessing or feature extraction since the raw signal is forwarded to a \gls{cnn}. Most computations are performed in the first two convolutional layers and are \gls{mac} operations due to the nature of convolution. The application is coded in C with 16-bit fixed-point arithmetic and includes a multicore version. 

\subsubsection{\Gls{cwm}}
The \gls{cwm}~\cite{j:zanetti2022} uses \gls{eeg} to monitor stress levels in 56-second windows. The most dominant kernel is feature extraction, which involves extracting statistical features in the time domain and power features in the frequency domain, accounting for more than \SI{65}{\percent} of the execution time. The application is coded in C using 32-bit \gls{fxp} arithmetic. The main operations are 32-bit multiplications with a 64-bit intermediate result followed by a shift (\gls{fxp} multiplications).

\subsubsection{\Gls{gcl}}
The \gls{gcl}~\cite{gestureclassifier} uses \gls{semg} to classify hand gestures in prosthetic arms within 0.2-second windows. The dominant kernel, the \gls{ica}, lies in the preprocessing step and involves matrix multiplications, translating into \gls{mac} operations. The application is coded in C using 32-bit \gls{fp} arithmetic and includes a multicore version.

\subsubsection{\Gls{coughdet}}
The \gls{coughdet}~\cite{orlandic_multimodal_2023} uses audio and \gls{imu} inputs to detect a coughing episode in 0.3-second windows. The dominant kernel is the \gls{mfcc}, which includes iterative \gls{fft} and transcendental computations (e.g., logarithm for the \gls{dct}). The application is coded in C using 32-bit \gls{fp} arithmetic.

\subsubsection{\Gls{ecl}}
The \gls{ecl}~\cite{Jose_bindi} classifies fear status using \gls{ppg}, \gls{gsr}, and \gls{st} in 10-second windows. The dominant kernel is the \gls{knn} inference, specifically, sorting the $n$-closest distances. The application is coded in C, and the \gls{knn} is implemented in 32-bit \gls{fp} arithmetic.

\subsubsection{\Gls{bpfree}}
The \gls{bpfree}~\cite{biobpfree} is a neural network training algorithm for resource-constrained devices. In BiomedBench, the algorithm trains the \gls{cnn} of \gls{seizdetcnn}. The main computations are vector-matrix and vector-vector multiplications arising from the forward and backward passes. The application is coded in C using 32-bit \gls{fp} arithmetic.

%% file: Experimental_setup.tex
\section{Experimental Setup}\label{sec:ExpSetUp}

% \usepackage{graphicx}

In this section, we discuss the methodology used to compare different \gls{mcu} architectures, the chosen platforms and their characteristics, and the measurement process. We start with an overview of our comparison methodology to clarify our workflow and enhance reproducibility in Section~\ref{sec:ExpSetup:Methodology}. We provide insight into the acquisition scenario and the programming of the \gls{mcu} operating modes and runtime in Section~\ref{sec:ExpSetup:Deployment}. We justify the choice and analyze the details of the evaluated boards in Section~\ref{sec:ExpSetup:Boards} and elaborate on the performance and energy measurement process in Section~\ref{sec:ExpSetup:Measurements}.  Furthermore, we assess the portability of the applications and the required software support in Section~\ref{sec:ExpSetUp:portability}. Lastly, we outline some deployment issues in specific applications and boards in Section~\ref{sec:ExpSetup:limitations}.

\subsection {Architectural comparison methodology}\label{sec:ExpSetup:Methodology}
The first step to performing insightful and reproducible comparisons among platforms is ensuring that all deployments follow the same assumptions and correctly utilize the platform operating modes and underlying hardware. More specifically, our implementations assumed that there is an external \gls{adc}, equipped with a buffer, capable of sending interrupts to the \gls{mcu} once the buffer is full. Then, our deployments fully utilized the sleep modes of the platform (i.e., retentive deep-sleep mode during idle, light-sleep mode during \gls{dma} acquisition) and operated in the optimal voltage and frequency levels for maximum energy efficiency during processing. The details of the acquisition scenario and the optimal strategies regarding the \gls{mcu} operating modes and state transitions are analyzed in Section~\ref{sec:ExpSetup:Deployment}.

Extracting impactful insights and holistically comparing various architectures necessitates a multi-dimensional evaluation. We selected three dimensions to compare various \gls{mcu} architectures: execution time, processing efficiency, and energy consumption. Execution time is a key metric to decide if a platform features enough computational capability to deal with the strict real-time demands of applications. Processing efficiency refers to the microarchitecture suitability to run biomedical kernels and is paramount to understanding which designs can lead to energy-efficient processing and, hence, sustainable wearable systems. Finally, measuring the energy consumption per application phase (i.e., idle, acquisition, and processing) is a direct measure of the overall energy efficiency of a platform and reveals the best low-power \gls{mcu} designs. We elaborate on how to evaluate and compare platforms across these three dimensions in Section~\ref{sec:ExpSetup:Measurements}.

\subsection {Sensor emulation, sleep modes and runtime orchestration}\label{sec:ExpSetup:Deployment}
For signal acquisition, we emulated the sensing device and the \gls{adc} functionality using an external board. We assume an external \gls{adc} with 768 bytes of RAM buffer similar to \cite{adc} since the selected \glspl{mcu} features an internal \glspl{adc} with insufficient bit precision. This buffer enables per-batch acquisition by transferring data only when the buffer is full. This batch acquisition scheme wakes up the core less frequently. We did not consider the energy of the sensing device and \gls{adc} in our experiments, as it is common for all platforms. Moreover, we assumed that the sensing device has no processing ability and that all the computations take part in the \gls{mcu}.

\begin{figure}[tp]
	\centering
	\includegraphics[width=.8\linewidth]{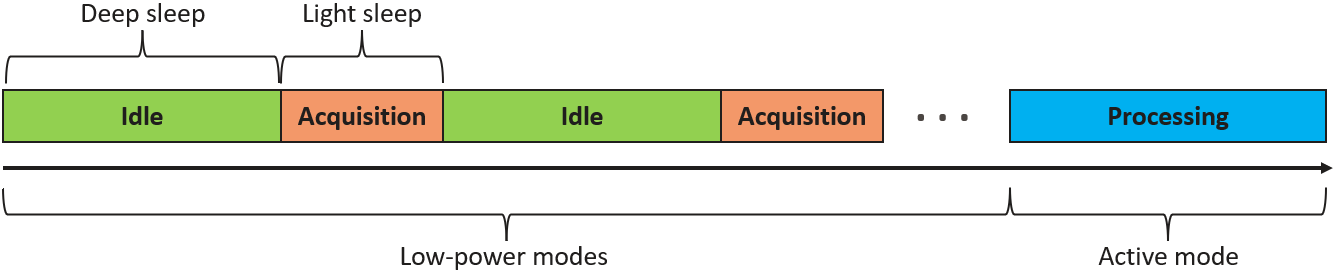}
	\caption{MCU operating phases under test.}
	\label{fig:mcu_operating_phases}
\end{figure}

~\figref{fig:mcu_operating_phases} shows the operating phases under evaluation. The \gls{mcu} is in a deep-sleep state when idle. After filling the buffer, the \gls{adc} sends an interrupt to awaken the \gls{mcu} that collects the signal samples. The data transfer uses the \gls{spi} protocol, with the evaluated board as master. For maximum energy efficiency, data movement to \gls{mcu} is executed through \gls{dma} while the core is in a light-sleep state. Upon acquisition of a full-window signal, the \gls{mcu} transitions to the active mode and executes the processing phase of the application.

During the idle period, we set the \gls{mcu} to deep-sleep mode with RAM retention to preserve the contents of the data needed for the next processing cycle. The \gls{mcu} must support a wake-up interrupt mechanism to switch to active mode when the data are ready. For the active phase of the processor, we select the lowest operating voltage that allows the processing frequency to meet the real-time constraints. Then, we configure the highest available frequency for this voltage, enabling the core to finish the computations and return to idle faster. This strategy yields the lowest energy consumption~\cite{hossein_frequency}, and we have validated it experimentally.

\begin{figure}[tp]
	\centering
	\includegraphics[width=.7\linewidth]{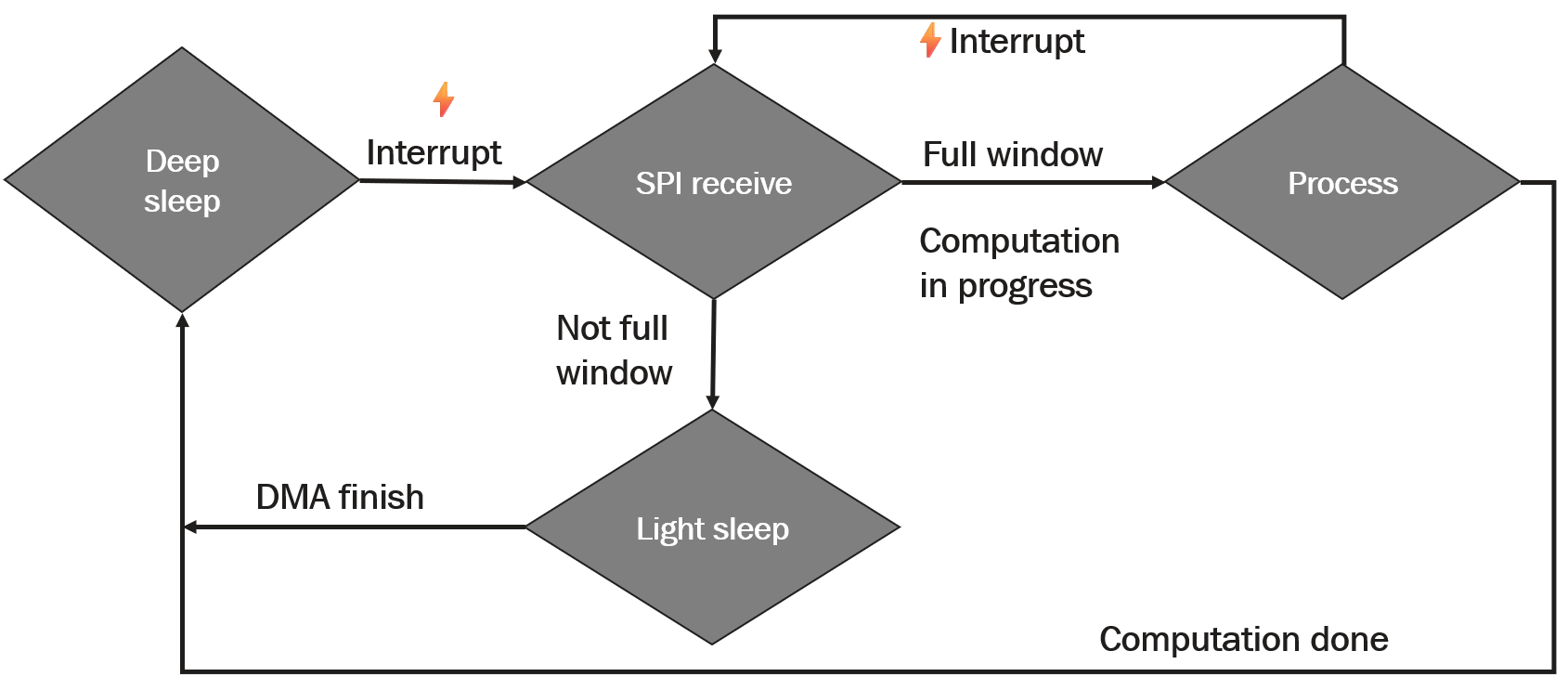}
	\caption{\gls{mcu} runtime state diagram.}
	\label{fig:mcu_state_diagram}      
\end{figure}

We orchestrated the application runtime and interchange between the \gls{mcu} states through interrupts and callback functions. Fig.~\ref{fig:mcu_state_diagram} depicts how the real-time execution is driven through different states. By default, the \gls{mcu} is in deep-sleep mode for maximum energy efficiency. Each time an interrupt is received from the external \gls{adc}, the \gls{mcu} wakes up from deep sleep and runs the external interrupt callback function. This triggers an \gls{spi} master receive, ideally served through the \gls{dma} while the core is in light sleep mode. When the \gls{dma} finishes the transfer, it raises an interrupt, and the core returns to deep sleep. Instead, if a full-window signal is captured, the core enters the active mode and executes the processing part before returning to deep-sleep mode. If an external interrupt arrives during processing, the callback function interrupts the processing to trigger an \gls{spi} receive, ensuring no part of the signal is lost. 

\begin{table}[tp]
\resizebox{.9\textwidth}{!}{%
\begin{tabular}{llllrrrr}
\hline
\textbf{\textbf{Board}} &
\textbf{\textbf{Manufacturer}} &
  \textbf{\textbf{MCU}} &
  \textbf{\textbf{Cores}} &
  \textbf{\textbf{FPU}} &
  \textbf{\textbf{RAM}} &
  \textbf{\textbf{Flash}} \\
 &
 &
   &
   &
   &
  \textbf{(\si{\kibi\byte})} &
  \textbf{(\si{\mega\byte})} \\ \hline
Raspberry Pi Pico &
Raspberry &
  RP2040 \cite{rp2040} &
  2x ARM Cortex-M0+ &
  No &
  264 &
  2 (off-chip)\\
Nucleo-L4R5ZI &
STMicroelectronics &
  STM32L4R5ZI \cite{stm32l4r5} &
  1x ARM Cortex-M4 &
  Yes &
  640 &
  2 (on-chip)\\
Ambiq Apollo3 &
Ambiq &
  Apollo3 Blue \cite{apollo3} &
  1x ARM Cortex-M4 &
  Yes &
  384 &
  1 (on-chip)\\
Gapuino &
GreenWaves &
  GAP8 \cite{gap8online} &
  1x CV32E40P (FC) &
  No &
  512 &
  2 (off-chip) \\
 &
 Technologies &
   &
  8x CV32E40P (Cluster) &
  Yes &
  64 & \\
GAP9EVK &
GreenWaves &
  GAP9 \cite{gap9} &
  1x CV32E40P (FC) &
  Yes &
  1564 &
  2 (off-chip) \\
 &
  Technologies &
   &
  9x CV32E40P (Cluster) &
  Yes &
  128 &\\ \hline
\end{tabular}%
}
\caption{Selected boards - Summary of basic features}
\label{Boards_table}
\vspace{-10pt}
\end{table}

\begin{table}[tp]
\resizebox{.8\textwidth}{!}{%
\begin{tabular}{lrrrrrr}
\hline
\textbf{\textbf{Board}} &
  \textbf{\textbf{Measuring}} &
  \textbf{\textbf{Voltage}} &
  \textbf{Optimal core} &
  \textbf{\textbf{Flash}} &
  \textbf{Sleep mode} &
  \textbf{DMA} \\
 &
 \textbf{\textbf{point}} &
   \textbf{(\si{\volt})} &
   \textbf{configuration} &
   \textbf{\textbf{used}} &
   \textbf{\textbf{used}} &
  \textbf{used} \\ \hline
Raspberry Pi Pico &
  Board &
  2.75 &
  \SI{0.90}{\volt}, \SI{133}{\mega\hertz} &
  Yes &
  Dormant mode &
  No \\
Nucleo-L4R5ZI&
  MCU &
  3.30 &
  \SI{1.20}{\volt}, \SI{120}{\mega\hertz} &
  Yes &
  Stop mode 2 &
  Yes \\
Ambiq Apollo3 &
  MCU &
  1.80 &
  \SI{0.70}{\volt}, \hspace{1.3 mm}\SI{48}{\mega\hertz} &
  Yes &
  Deep sleep &
  No \\
Gapuino &
  MCU &
  2.80 &
  \SI{0.80}{\volt}, \SI{150}{\mega\hertz} &
  No &
  - &
  Yes \\
    &
    &
  &
  \SI{0.80}{\volt}, \SI{150}{\mega\hertz} &
  &
   &\\
GAP9EVK &
  MCU &
  2.80 &
  \SI{0.65}{\volt}, \SI{240}{\mega\hertz} &
  No &
  Retentive &
  Yes \\
 &
   &
   &
  \SI{0.65}{\volt}, \SI{240}{\mega\hertz} &
  &
   Deep sleep &
   \\ \hline
\end{tabular}%
}
\caption{Configuration used for each board}
\label{Boards_table_setup}
\vspace{-10pt}
\end{table}

\subsection {Evaluated low-power boards}\label{sec:ExpSetup:Boards}
Platform selection is essential to make meaningful comparisons and to gain insight into the critical hardware characteristics that can improve performance and energy efficiency. Architectural variability is essential to achieving insightful comparisons. Our selection of boards was based on variability regarding \glspl{isa} (i.e., ARM and RISC-V), processors with the same \gls{isa} (i.e., Arm Cortex-M0+ and Arm Cortex-M4; CV32E40P GAP8, and CV32E40P GAP9), and \gls{mcu} designs (i.e., STM32, GAP, Apollo, RP2040). 

We selected five popular commercial low-power boards featuring five different \glspl{mcu} and four different processors, including two different \glspl{isa}. The selected boards are a Nucleo-L4R5ZI~\cite{nucleol4r5} from the \gls{ulp} line of ST Microelectronics, an Ambiq Apollo3 Blue~\cite{apollo3}, a Raspberry Pi Pico~\cite{picoHome}, a Gapuino v1.1~\cite{gapuino} and a GAP9EVK~\cite{gap9store} from GreenWaves Technologies. \tblref{Boards_table} summarizes the core characteristics of the boards, describing the architecture and storage capabilities of each \gls{mcu}. All boards include deep-sleep modes, \gls{spi}, and \gls{dma} support to manage data movements from \gls{spi} to RAM.

We summarize the main features of the measurements per board in \tblref{Boards_table_setup}. The first two columns refer to the measuring point and its voltage level. The third column describes the optimal core configuration that yields the best energy efficiency. The final three columns refer to the flash usage, the sleep mode used, and the \gls{dma} utilization for signal acquisition.

A more detailed description of each board follows.

\subsubsection{Raspberry Pi Pico}
This platform includes two identical cores. All code and data are stored in the RAM for maximum efficiency, except for cases of insufficient capacity. In these cases, static data are stored in the flash, inducing negligible runtime overhead. We have ported the applications using the Pico \gls{sdk}~\cite{pico-sdk} and ARM compiler. Since \gls{sdk} does not provide sufficient support, we could not employ the \gls{dma} unit or shut down individual memory banks to minimize energy consumption.

\subsubsection{Nucleo-L4R5ZI}
The code is stored in flash memory with zero overhead for instruction fetching, while the rest of the data resides in SRAM. We use \gls{dma} to manage data acquisition while configuring the core to the lowest sleep mode that allows \gls{dma} to operate. We have ported the applications using the CubeIDE \cite{stm32cubeide-sdk} and the ARM GCC compiler. 

\subsubsection{Ambiq Apollo3}
In sleep mode, we power down the unneeded RAM banks and retain only the needed ones. The code is stored in a flash with zero overhead in instruction fetching, whereas the rest of the data resides in SRAM. In \gls{cnn}, the static data reside in the flash. We also use TurboSPOT mode with the core running at \SI{96} {\mega\hertz} to meet the time constraints of \gls{seizdetcnn}, \gls{coughdet}, and \gls{gcl}. We have ported the applications using the Apollo Ambiq Suite \gls{sdk} \cite{apollo3} and the ARM GCC compiler. Due to insufficient support from the \gls{sdk}, we cannot use the \gls{dma} for \gls{spi} acquisition.

\subsubsection{Gapuino}
All the RISC-V cores in Gapuino include microarchitectural \gls{dsp} extensions~\cite{riscvdspextensions}. All data and code reside in the SRAM for maximum performance. Despite our best efforts, we did not succeed in using the deep-sleep mode or memory bank management capabilities of this board. Therefore, we put the core in active waiting for the idle period and scale down the frequency to the minimum of \SI{32}{\kilo\hertz} while keeping the cluster powered off. We have ported the applications using the GAP \gls{sdk} \cite{gap8-sdk} and the RISC-V GNU compiler toolchain. Programming the cluster necessitates the use of PMSIS \gls{api}.

To achieve maximum performance on this board, one should move the data to the L1 memory and run the computations on the cluster. This is the only way to ensure single-cycle memory access and comes with the cost of moving the data from the L2 memory to the L1, which has a limited capacity of \SI{64}{\kibi\byte}. The overhead of data transfers is usually low since it is served through a fast \gls{dma}. However, our experiments with single-core applications suggest that running the computations on the main core is more energy efficient and avoids the energy overhead of powering up the cluster.

\subsubsection{GAP9EVK}
This board has an architecture similar to that of Gapuino. In addition, the GAP9 cores are enhanced with more custom instructions and \gls{fpu}, higher operating frequencies, lower technology nodes (\SI{22}{\nano\meter} vs. \SI{55}{\nano\meter}), one additional CPU, and more memory. The retentive sleep mode on this board enables the fine-grain power management of different banks. We have ported the applications using a private version of GAP \gls{sdk}, licensed only under \gls{nda}, and the RISC-V GNU compiler toolchain. The programming of the cluster requires the use of the PMSIS \gls{api}. Like in Gapuino, data transfers to L1 memory are essential for maximum performance. However, single-core applications are more energy efficient running on the main core while keeping the cluster powered down.

We do not employ hardware accelerators (e.g., GAP neural network accelerators) since we focus on comparing processor performance. The use of accelerators is outside the scope of our experiments.

\subsection{Measurements}\label{sec:ExpSetup:Measurements}
\subsubsection {Instruction mix}
The instruction mix showcases the percentage of each instruction group issued while executing the processing part of each application. This kind of analysis assists in understanding the applications' computational profile at a higher level. We obtained the instruction mix by dumping the dynamic assembly trace with GDB running on an ARM Cortex-M4 processor. 

For a more straightforward interpretation, we conceptually grouped instructions by their purpose. The \emph{LOAD} and \emph{STORE} instruction groups showcase the memory access profile of the application, eventually hinting at the need for a vertical hierarchy. The \emph{BRANCH} group includes conditional execution (i.e., branches and conditional assignments). The \emph{INTEGER} and \emph{SHIFT} groups contain the main computational part of the code, with arithmetical instructions and shifts. Then, in the \emph{FLOAT} group, we find all the \gls{fp} instructions (i.e., multiplications, additions, comparisons), highlighting the need for \gls{fp} hardware support. Finally, all the other possible instructions are unified under the \emph{OTHERS} group, containing mainly data movement instructions and bit-wise operations.

\subsubsection {Execution time}
The first criterion when evaluating a platform is its ability to execute applications within real-time constraints. A platform is eligible for real-time patient monitoring only if it completes an application processing phase within its window duration. Not all platforms manage to meet the real-time demands of all applications. This can happen due to insufficient computational capability, especially when the application window is short.

We measured the processing time using the Otii Arc~\cite{otii} and GPIO signals. The Otii Arc provides enough resolution to accurately measure the execution time and evaluate the suitability of a platform when dealing with real-time constraints of applications. We highlight that the reported processing times regard the most energy-efficient configurations of platforms and cannot serve as a raw speed comparison among platforms. Moreover, multicore executions are mixed with single-core executions. When a platform fails to complete processing within the application window, we report the execution speed for the maximum frequency available.

\subsubsection {Processing efficiency}
We evaluated the processing efficiency of different processors using the number of processing cycles as a metric. This metric is independent of the differences in operating frequencies between platforms and can help directly compare the processing capabilities of different platforms. To neutralize the effect of memory while evaluating the processors, all application executions must have the same memory access latency on all platforms. Apollo3 Blue, Raspberry Pi Pico, and STM32L4R5ZI feature single-cycle access to main memory. To ensure single-cycle access for the CV32E40P cores on Gapuino and GAP9EVK, we run the code on the cluster using the L1 memory and do not consider the cycles spent moving data to L1.

To measure the processing cycles, we employed two different methods. First, we utilized the built-in hardware performance counters when the processor supports them (i.e., Arm Cortex-M4, CV32E40P). Second, we measured the processing time using the Otii Arc and GPIO signals and divided it by the operating frequency. We have confirmed that both methods yield the same results on processors supporting the first option.

On top of measuring the cycles of the complete application, we executed a series of custom experiments to evaluate the processors' efficiency running specific instructions (e.g., 32-bit \gls{fxp} multiplications, shifts, \glspl{mac}, branches). We built custom loops with the instructions of interest and calculated the cycles per instruction using the total cycles and disassembly of the loop. Our results are consistent with the cycles per instruction reported in the datasheets of each processor. This information is vital to understanding the strengths and weaknesses of each processor and explaining the performance discrepancies when running the benchmark applications. 

\subsubsection {Energy}\label{sec:ExpSetup:Energy}
We use manufacturer-provided evaluation boards to monitor each \gls{mcu} energy consumption during the execution of the application. These boards offer built-in circuits to measure current and voltage directly from the \gls{mcu}. All energy measurements were taken at the power supply entry point of the \gls{ic}. The reported energy numbers include the energy drawn by the input-to-core step-down voltage converter embodied in the integrated circuit. Most of the evaluated platforms integrate an inefficient \gls{ldo} that dissipates the unused energy during the power conversion stage---only the Apollo3 Blue has a more efficient DC-DC buck converter.

Gapuino provides an internal \SI{1}{\ohm} resistor to measure the current drawn by the \gls{mcu} through the voltage drop. Nucleo-L4R5ZI, Apollo3 Blue, and GAP9EVK have open circuit pins where we placed a \SI{1}{\ohm} resistor to measure the current with the same mechanism. The maximum current drawn by \glspl{mcu} was lower than \SI{30}{\milli\ampere} while the input voltage drop was within the operating range of each device. We have measured the total board consumption for Raspberry Pi Pico since it has no test points for the \gls{mcu}.

We have selected the Otii Arc supplied by Qoitech to take measurements on each board. This device samples at \SI{4}{\kilo\hertz}, which allowed us to obtain an energy profile over time and extract the energy and execution time per phase. However, due to the limited $\pm$\SI{10}{\micro\ampere} precision of the Otii device, we used the Fluke 8846A~multimeter~\cite{fluke} to measure the average current of Nucleo-L4R5ZI and Apollo3 Blue in deep-sleep mode. With this device, we can achieve a precision of \SI{0.03}{\micro\ampere} when used in the current measurement mode, which operates with a \SI{100}{\ohm} resistance. 

\subsection{Software support and portability}\label{sec:ExpSetUp:portability}
All the applications are written in C/C++ and compiled with -O3 optimizations. Since modern compilers are highly optimized and exploit all \gls{isa} extensions present (e.g., \gls{dsp} instructions, hardware loops), we ensure full utilization of underlying hardware and a fair comparison across platforms. For more details on the code and compilation flags, check out the open-source GitHub repository\footnote{https://github.com/esl-epfl/biomedbench}.

We use the FreeRTOS API to manage dynamic memory allocation and ensure the portability of the code across the platforms since all selected boards feature FreeRTOS support. However, FreeRTOS does not suffice to manage deep-sleep modes, communication with peripherals, and multicore implementations. Instead, the \gls{hal} of the board \gls{sdk} provides the interface to program these functionalities. In particular, data acquisition requires \gls{spi} peripheral communication, sleep mode transitions require the configuration of the \gls{pmu} with selective retention for SRAM banks, and multicore implementations require a multicore \gls{api}. Only the \gls{hal} supports programming these functionalities, as FreeRTOS does not provide a standard interface to achieve these objectives. \textbf{\textit{Takeaway~1: To achieve fully portable applications in the wearable domain, FreeRTOS should be extended and ported across platforms to provide standard support for fine-grained sleep modes, communication with peripherals, and multicore execution.}}

\subsection{Deployment limitations}\label{sec:ExpSetup:limitations}
\subsubsection{Instruction mix}
For the instruction mix, we have obtained results for six of the eight applications.\footnote{The instrumented versions of \gls{bpfree} and \gls{coughdet} continued to crash the GDB. Therefore, we limit our study to the rest of the applications for which we managed to obtain a complete execution trace}
% The GDB crashed several times during evaluation for \gls{bpfree} and \gls{coughdet}. Considering that we managed to obtain a decent high-level view through the instruction profile of the six applications, we decided to abandon the efforts for these two applications.
\subsubsection{Deployment of \texorpdfstring{\gls{coughdet}}{CoughDet}}
\gls{coughdet} requires \SI{568}{\kibi\byte} for storing static parameters (mainly for the \gls{lut} and the \gls{rf}). We did not deploy \gls{coughdet} on Gapuino since there is insufficient on-chip memory. Moreover, Gapuino features no on-chip flash; hence, using the external flash would compromise the performance and energy results. Using the on-chip flash, we overcame this issue on Apollo3 Blue, Raspberry Pi Pico, and Nucleo-L4R5ZI. GAP9EVK features sufficient on-chip RAM.

\subsubsection{Deployment of \texorpdfstring{\gls{bpfree}}{Bio-BPfree}}
\gls{bpfree} has the largest memory footprint of all applications as illustrated in \tblref{App_Metrics_Table}. To port \gls{bpfree} to our selected platforms, it is necessary to utilize the flash and constantly move data in and out of the RAM. To avoid the complexity of such an implementation and ensure a fair performance comparison among the selected platforms, we apply \gls{bpfree} to just one convolutional layer of the \gls{cnn} network and scale down the dimensions of the convolution. We use an input of length $64$ with $32$ channels and $64$ filters. Scaling down the dimensions does not change the computational profile of the application and allows us to fit all non-static data in the RAM during runtime.

%% file: Experimental_results.tex
\section{Experimental results}\label{sec:ExpRes}

This section analyzes the results of our experiments running BiomedBench on SoA platforms. In Section~\ref{sec:ExpRes:Instructions}, we present the instruction mix of the applications running on an Arm Cortex-M4. In Section~\ref{sec:ExpRes:Time}, we report the applications' execution time for each platform and showcase if a platform is eligible to tackle real-time applications' limitations. In Section~\ref{sec:ExpRes:Transcendental}, we highlight the necessity to employ an \gls{lut} for transcendental computations due to the lack of advanced computational units in low-power architectures. In Section~\ref{sec:ExpRes:Cycles}, we compare the execution cycles per processor for different applications and analyze the microarchitectural differences among processors and their impact on performance. In Section~\ref{sec:ExpRes:Energy}, we report the platforms' energy consumption per phase for different applications, analyzing their strengths and weaknesses. Finally, we summarize the key takeaways of this section to provide hints for future software and hardware designers in Section~\ref{sec:ExpRes:Summary}.

 \begin{figure}[tp]
	\centering
	\includegraphics[width=.65\linewidth]{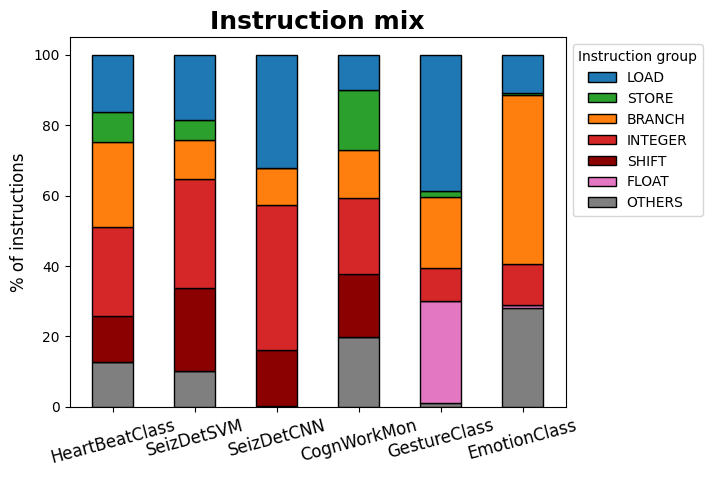}
	\caption{Instruction mix of BiomedBench applications running on ARM Cortex M4.}
	\label{fig:instruction_mix}
\end{figure}

\subsection{Instruction mix}\label{sec:ExpRes:Instructions}
~\figref{fig:instruction_mix} shows the distribution of different kinds of instructions executed during the full application cycle of six benchmark applications on Arm Cortex-M4. The instruction mix can provide a high-level view of the computation profile of each application and, possibly, hint at microarchitectural features that will efficiently handle specific workloads.

The instruction mix results present interesting variability among the six applications. Regarding memory interaction, load instructions are prominent in \gls{seizdetcnn} (convolution) and \gls{gcl} (matrix multiplication), significantly outnumbering the store instructions. Branch instructions appear in very high proportion in \gls{ecl} (\gls{knn} inference) and \gls{hcl} (\gls{mf}), while they are less frequent in the rest of the applications. Integer arithmetic represents a significant percentage of the total instructions in \gls{fxp} applications, mainly in \gls{seizdetsvm} and \gls{seizdetcnn}, while it is less relevant in \gls{fp} applications. \gls{fxp} applications also include a noticeable amount of shift instructions -- linked to \gls{fxp} multiplications and format conversions. \gls{seizdetsvm} and \gls{cwm} are characteristic examples of increased shift operations. They use different fixed-point representations in each module to avoid losing precision, thus using shifts to convert between different fixed-point representations. Finally, \gls{gcl} employs a significant amount of floating-point operations, requiring a strong \gls{fpu} for high performance. \textbf{\textit{Takeaway~2: \Acrfull{fxp} applications feature increased shift operations and benefit from the presence of a barrel shifter for efficient processing.}}

\subsection{Execution time}\label{sec:ExpRes:Time}

\begin{table}[tp]
\begin{center}
\begin{tabular}{@{\hskip0.05cm}l|@{\hskip0.1cm}r|@{\hskip0.1cm}@{\hskip0.1cm}r@{\hskip0.1cm}r@{\hskip0.1cm}r@{\hskip0.05cm}r@{\hskip0.05cm}r@{\hskip0.05cm}r} 
 \toprule
 & \multirow{2}{*}{\textbf{Window (s)}} &
 \multicolumn{5}{c}{
        \textbf{Execution time (s)}
 } \\
 &  &
 \multicolumn{1}{@{\hskip0.1cm}c}{\textbf{Apollo3 Blue}} & \multicolumn{1}{@{\hskip0.1cm}c}{\textbf{STM32L4R5ZI}} & \multicolumn{1}{@{\hskip0.1cm}c}{\textbf{GAP8}} & \multicolumn{1}{@{\hskip0.1cm}c}{\textbf{GAP9}} &
 \multicolumn{1}{@{\hskip0.1cm}c}{\textbf{RP2040}}\\

 \midrule
\gls{hcl} & \textbf{15.0} & 0.115 & 0.046 & {\hskip0.05cm} (M8) 0.026  & {\hskip0.05cm} (M9) 0.012 & {\hskip0.05cm} (M2) 0.045\\ 
\gls{seizdetsvm} & \textbf{60.0} & 0.058 & 0.019 &  {\hskip0.05cm}(M8) 0.009 & {\hskip0.05cm}(M9) 0.002  & {\hskip0.05cm}(M2) 0.022 \\
\gls{seizdetcnn} & \textbf{4.0} & 2.341 & 2.010 & {\hskip0.05cm}(M8) 0.232  & {\hskip0.05cm}(M9) 0.070  & {\hskip0.05cm}(M2) 1.451 \\
\gls{cwm} & \textbf{56.0} & 2.930 & 1.149 & 1.501 & 0.383 & 2.600\\
\gls{gcl} & \textbf{0.2} & \textcolor{red}{0.239} & 0.191 & {\hskip0.05cm}(M8) \textcolor{red}{1.015}  & {\hskip0.05cm}(M9) 0.011  & {\hskip0.05cm}(M2) \textcolor{red}{2.529} \\
\gls{coughdet} & \textbf{0.3} & 0.208 & 0.086 & - & 0.038  & \textcolor{red}{1.125}\\
\gls{ecl} & \textbf{10} & 0.054 & 0.021 & 0.095 & 0.007 & 0.115\\
\gls{bpfree} & - & 13.824 & 5.521 & 123.027 & 2.652 & 125.791\\

\bottomrule
\end{tabular}%
\end{center}
\caption{Execution time in \si{seconds} per \gls{mcu} and application. Multicore execution times are reported whenever possible, denoted as M2 for Pico, M8 for GAP8, and M9 for GAP9 (i.e., 2, 8, and 9 cores, respectively). We have selected the frequency and voltage combination for each platform that results in the lowest energy consumption for each application window; therefore, this table does not serve as a comparison of the lowest possible execution times. For example, the multicore implementation on the RP2040 seems faster than that on the single-core Apollo3 Blue for \gls{seizdetcnn} (\SI{1.4}{\second} and \SI{2.3}{\second}, respectively); however, we selected an operating frequency of \SI{96}{\mega\hertz} for the Apollo3 Blue and of \SI{133}{\mega\hertz} for the RP2040. The lowest frequency of the Apollo3 Blue still allows it to meet deadlines while minimizing its energy consumption. For the red cases, we report the execution time for the maximum frequency available.}
\label{TimeperBoard}
\vspace{-18pt}
\end{table}

\tblref{TimeperBoard} summarizes the processing time required for each platform and the operating window duration of each application. This table showcases the raw speed potential of the platforms since it reports the execution time at maximum frequency. In red, we highlight the execution times larger than the operating window, in which case the platform cannot tackle the real-time demands of the application. 

Only two of the eight applications challenge the platforms under test. \gls{gcl} and \gls{coughdet} feature sub-second windows and \gls{fp} computations. Gapuino and Raspberry Pi Pico fail to run the processing phases of these applications within the given time frame due to the lack of \glspl{fpu}. Apollo3 Blue, operating at \SI{96}{\mega\hertz}, fails to support the real-time execution of \gls{gcl}. In contrast, Nucleo-L4R5ZI, featuring the same processor but operating at \SI{120}{\mega\hertz}, completes \gls{gcl} within the time frame.  

GAP9 dominates its competitors in performing fast computations thanks to its highest operating frequency of \SI{240}{\mega\hertz} and the presence of the multicore cluster. GAP9 completes the processing part of the applications from $1.9\times$ (\gls{cwm}) to $28.7\times$ (\gls{seizdetcnn}) faster than STM32L4R5ZI. In addition, GAP9 outperforms its predecessor GAP8, running applications from $3.3\times$ (\gls{cwm}) to $92.3\times$ (\gls{seizdetcnn}) faster thanks to its \gls{isa} extensions, increased operating frequency, and \gls{fpu}. In general, GAP9 has the highest computational ability and can comfortably deal with the tightest real-time constraints (that is, \gls{gcl}, \gls{coughdet}). \textbf{\textit{Takeaway~3: GAP9 has the highest computational ability thanks to its high operating frequency, multicore cluster, \gls{fpu}, and \gls{dsp} \gls{isa} extensions and is the best choice for developing and deploying applications with tight real-time constraints.}}

\begin{table}[tp]
\resizebox{0.65\textwidth}{!}{%
\begin{tabular}{@{}lrrrr@{}}
\toprule
 &
  \textbf{\begin{tabular}[c]{@{}r@{}}CV32E40P \\ GAP8\end{tabular}} &
  \textbf{\begin{tabular}[c]{@{}r@{}}CV32E40P \\ GAP9\end{tabular}} &
  \textbf{\begin{tabular}[c]{@{}r@{}}ARM \\ Cortex-M0+\end{tabular}} &
  \textbf{\begin{tabular}[c]{@{}r@{}}ARM \\ Cortex-M4\end{tabular}} \\ \midrule
\gls{hcl}           & 5.1 & 5.1 & 11.6 & 7.4 \\ 
\gls{seizdetsvm}    & 2.8 & 2.5 & 4.3 & 2.3 \\
\gls{seizdetcnn}    & 160.0 & 160.0 & 283.0 & 240.0 \\
\gls{cwm}           & 165.0 & 92.0 & 346.0 & 138.0 \\
\gls{gcl}           & 635.8 & 20.2 & 571.6 & 22.9 \\
\gls{coughdet}      & - & 9.1 & 149.7 & 9.9 \\
\gls{ecl}           & 12.7 & 1.6 & 15.3 & 2.5 \\
\gls{bpfree}        & 18450.0 & 633.0 & 16758.0 & 662.0 \\
\bottomrule
\end{tabular}%
}
\caption{Millions of processing cycles per core}
\label{Cycles_per_cpu}
\vspace{-10pt}
\end{table}

\subsection{Transcendental computations}\label{sec:ExpRes:Transcendental}

In biomedical applications, there is often the need to perform computations involving transcendental functions, such as cosine and sine. These are usually implemented in libraries through time-hungry algorithms, so encoding the results in a \gls{lut} is a solution that must be considered. This approach can save significant computations at the cost of additional memory footprint. To better explain this, we focus on the \acrfull{dct}. This algorithm is employed in the \gls{coughdet} application to compute the \gls{mfcc} features of an audio signal. The transform is applied to a matrix with a linear transform function, \emph{\_dct\_linear()}, which is called ten times (once per each column of the matrix). Inside the linear transform, the cosine must be evaluated 16384 times, leading to 163840 cosine computations per data window.

\begin{table}[tp]
\resizebox{.5\textwidth}{!}{%
    \begin{tabular}{l r r}
    \toprule
    \multirow{2}{*}{\textbf{Method}} &
    \multicolumn{2}{c}{
        \textbf{Exec. time (ms)}
    }\\
    & \textbf{Arm Cortex-M4} & \textbf{CV32E40P GAP9}\\ 
    \midrule
    Look-Up-Table & 1.46 & 0.28 \\
    \emph{cosf()} from math.h & 184.66 & 80.56 \\
    \emph{cos()} from math.h & 507.57 & 547.82 \\ 
    \hline
    \end{tabular}%
}
\caption{Execution times in milliseconds of the linear \acrfull{dct} function on Arm Cortex-M4}
\label{DCT exec times}
\vspace{-10pt}
\end{table}
To show the importance of implementing a \gls{lut} for the cosine, we compare the execution time of the \emph{\_dct\_linear()} function for three different computation methods. In particular, we highlight the difference between reading the pre-computed value from memory (via the \gls{lut}) and computing it through the \emph{cosf()} and \emph{cos()} functions from the \emph{math.h} library (single and double floating-point precision, respectively). The \gls{lut} method is computation-free since it stores all the values needed (the cosine arguments are not data-dependent), so it only retrieves the resulting value from memory. The memory overhead of the \gls{lut} is \SI{64}{\kibi\byte}.

\tblref{DCT exec times} reports the results of such experiments running on the STM32L4R5ZI and CV32E40P GAP9. Comparing the \gls{lut} method with the cosf() function, we obtained a speedup of $126.47\times$ and $286.7\times$ for the two boards. GAP9, which runs at double the frequency of STM32L4R5ZI, yields a higher speedup since the iterative loop for the \gls{dct} can be significantly simplified using the hardware loop capabilities of this platform. Moreover, in this specific case, it is possible to save up to \SI{182.2}{\milli\second} (for the STM32L4R5ZI), which is \SI{61.06}{\percent} of the total window (\SI{0.3}{\second}). This emphasizes how important it can be to avoid transcendental computations paying the price of memory occupancy. \textbf{\textit{Takeaway~4: Exploiting a \acrfull{lut} to avoid transcendental computations can yield more than \boldmath$100\times$ speedup gains with moderate memory overhead when deploying biomedical kernels on low-power microcontrollers.}}

\subsection{Processing efficiency}\label{sec:ExpRes:Cycles}
In \tblref{Cycles_per_cpu}, we demonstrate the total processing cycles per application and processor. Our results highlight interesting variability across different applications and processor architectures.

\subsubsection{\gls{hcl}}\label{sec:ExpRes:Perf:HCL}
CV32E40P outperforms ARM, requiring 31\% fewer cycles than ARM Cortex-M4 and 47\% fewer cycles than ARM Cortex-M0+. The main kernel of this application is the \gls{mf} implemented through a queue involving multiple conditional data movements and minimum/maximum searches. The CV32E40P cores benefit from min/max instructions that execute in only one cycle and a larger register file of 32 registers against the 16 registers of the ARM processors. When inspecting the disassembly, it is obvious that this application stresses the register file and necessitates numerous extra stack pushes and pops in ARM. ARM Cortex-M0+ suffers more from this phenomenon as it has a weaker instruction set than ARM Cortex-M4 and has to allocate more registers for the same operations. \textbf{\textit{Takeaway~5: Limiting the register file to 16 registers may lead to a significant performance drop due to increased stack push/pop operations. This increase in memory accesses negates any benefit of using a smaller register file. Arm Cortex-M processors experience this issue when running the \acrfull{mf} kernel implemented with a queue.}}

\subsubsection{\gls{seizdetsvm} - \gls{cwm}}\label{sec:ExpRes:Perf:SVM-CWM}
Both applications include \gls{fxp} multiplications of 32-bit integers with a 64-bit result followed by a shift. These multiplications take one cycle in ARM Cortex-M4 and CV32E40P GAP9, five cycles in CV32E40P GAP8, and more than 15 cycles in ARM Cortex-M0+. Moreover, the shift of the 64-bit intermediate involves two shifts of 32-bit registers and an "OR" operation between them. The 64-bit register shift takes two cycles in Arm Cortex-M4 and three in the other processors. Overall, an \gls{fxp} multiplication takes three cycles on Arm Cortex-M4, four cycles on CV32E40P GAP9, eight cycles on CV32E40P GAP8, and more than 15 cycles on Arm Cortex-M0+. As a result, ARM Cortex-M4 and CV32E40P GAP9 are the most efficient microarchitectures for these types of applications. \textbf{\textit{Takeaway~6: The ability of a processor to efficiently perform 32-bit fixed-point multiplications with a 64-bit intermediate result is essential to achieve maximum performance in 32-bit fixed-point biomedical applications.}}

In \gls{cwm}, CV32E40P GAP9 requires 33\% fewer cycles than Arm Cortex-M4 despite having comparable performance when dealing with 32-bit \gls{fxp} multiplications. This performance difference is mainly attributed to the fact that this application includes many loops with a small number of operations in the loop body (i.e., vector operations, vector copies, change of \gls{fxp} representation). The zero-overhead hardware loops of CV32E40P GAP9 provide significant performance gains in loops with a small number of operations in the body. On top of that, Arm processors struggle to execute the biquad filter due to their reduced 16-bit register file, forcing them to interact with the stack often. CV32E40P GAP8 requires 78\% more cycles than CV32E40P GAP9 due to the large amount of 32-bit multiplications. For the same reason, Arm Cortex-M0+ requires 149\% more cycles than Arm Cortex-M4.

In \gls{seizdetsvm}, apart from the 32-bit \gls{fxp} multiplications, there is a significant number of 32-bit \gls{fxp} divisions. 32-bit \gls{fxp} divisions involve the division of 64-bit operands, which is not supported by the hardware and is emulated in software. Arm Cortex-M4 outperforms CV32E40P GAP9 by 8\%, thanks to a superior implementation of the 64-bit division emulation. CV32E40P GAP9 requires 11\% fewer cycles than CV32E40P GAP8 since it completes 32-bit \gls{fxp} multiplications in half the cycles. Finally, Arm Cortex-M0+ requires 87\% more cycles than Arm Cortex-M4 due to the slower 32-bit \gls{fxp} multiplications.

\subsubsection{\gls{seizdetcnn}}\label{sec:ExpRes:Perf:CNN}
CV32E40P processors significantly outperform Arm, requiring 33\% fewer cycles than ARM Cortex-M4 and 50\% fewer cycles than ARM Cortex-M0+. This application consists of \gls{mac} operations on 16-bit \gls{fxp} numbers with 32-bit results followed by a shift. These operations take one cycle in CV32E40P, two cycles in the ARM Cortex-M4 (one for the \gls{mac} and one for the shift), and three cycles in the ARM Cortex-M0+ (two for the \gls{mac} and one for the shift). Moreover, this application includes quadruple nested loops with a predefined number of iterations and some conditional branches. Considering that branching performance is comparable in all architectures, CV32E40P outperforms the other cores by exploiting the \gls{dsp} extensions, including one-cycle \gls{mac}/shift, zero overhead hardware loops, and post-load/store pointer increment in one cycle. ARM Cortex-M4 also features some \gls{dsp} instructions, including one-cycle \gls{mac} operations and post-load/store pointer increment, while ARM Cortex-M0+ does not have \gls{dsp} specialized instructions. \textbf{\textit{Takeaway~7: \Acrfull{dsp} microarchitectural extensions and hardware loops can provide great advantages when dealing with computationally intensive biomedical applications.}}

\subsubsection{\gls{gcl} - \gls{coughdet} - \gls{bpfree}}
CV32E40P GAP9 and ARM Cortex-M4 outperform the other two processors due to their \glspl{fpu}. This is natural since these applications mainly feature multiplications in 32-bit \gls{fp} arithmetic. It is interesting to note that CV32E40P GAP9 outperforms ARM Cortex-M4 in \gls{fpu} performance since it completes a 32-bit float \gls{mac} in one cycle versus three cycles and also performs the loads and stores in one cycle instead of two. Furthermore, it is clear that CV32E40P GAP8 and ARM Cortex-M0+, which lack an \gls{fpu}, are poor choices for applications that involve \gls{fp} operations and cannot be easily converted to \gls{fxp} implementations without accuracy drop. 

\subsubsection{\gls{ecl}} \label{sec:ExpRes:Perf:ECL}
CV32E40P GAP9 dominates ARM Cortex-M4 in this floating-point application, requiring 36\% fewer cycles. The dominant kernel of this application is sorting the Euclidean distances, featuring a nested loop with \gls{fp} comparisons and conditional assignments. CV32E40P GAP9 executes the conditional assignment of minimum distance and index using conditional branching, taking three cycles if a new minimum is found (branch not taken) or four cycles (branch taken). On the contrary, ARM Cortex-M4 uses specialized conditional assignment instructions instead of conditional jumps and executes the same workload in six cycles. Moreover, CV32E40P GAP9 is boosted by zero-overhead hardware loops (zero vs. three cycles), leading to a significant performance difference. On the other hand, CV32E40P GAP8 and ARM Cortex M0+ struggle with this application due to the lack of \glspl{fpu} for the \gls{fp} comparisons.

\begin{figure}[tp]
	\centering
	\begin{subfigure}{\textwidth}
        \centering
		\includegraphics[width=1.0\textwidth, height=.6\textwidth]{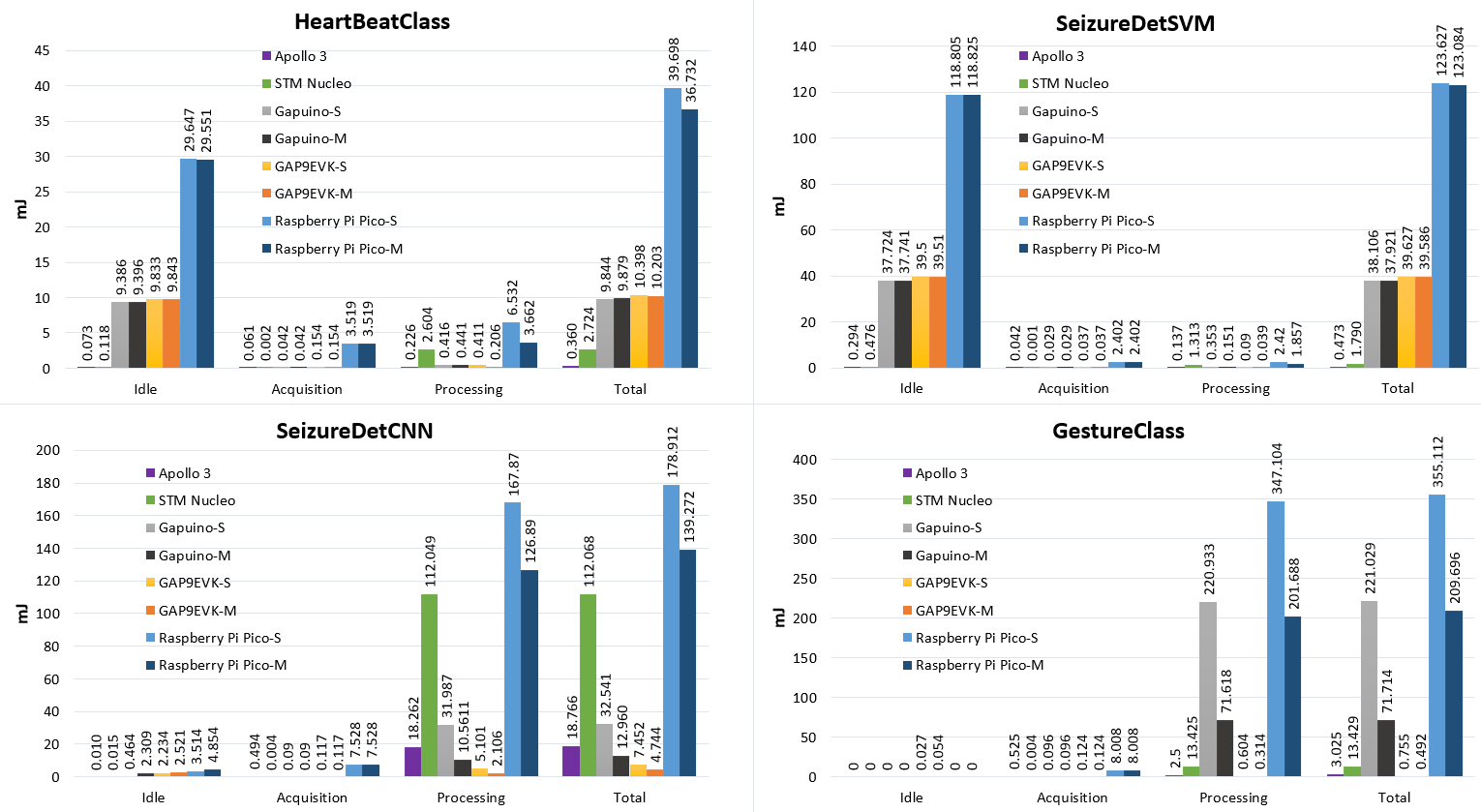}
		\caption{Applications with multicore option}
	\end{subfigure}
%%%%%%%%%%%%%%
	\begin{subfigure}{\textwidth}
        \centering
		\includegraphics[width=0.8\textwidth, height=.5\textwidth]{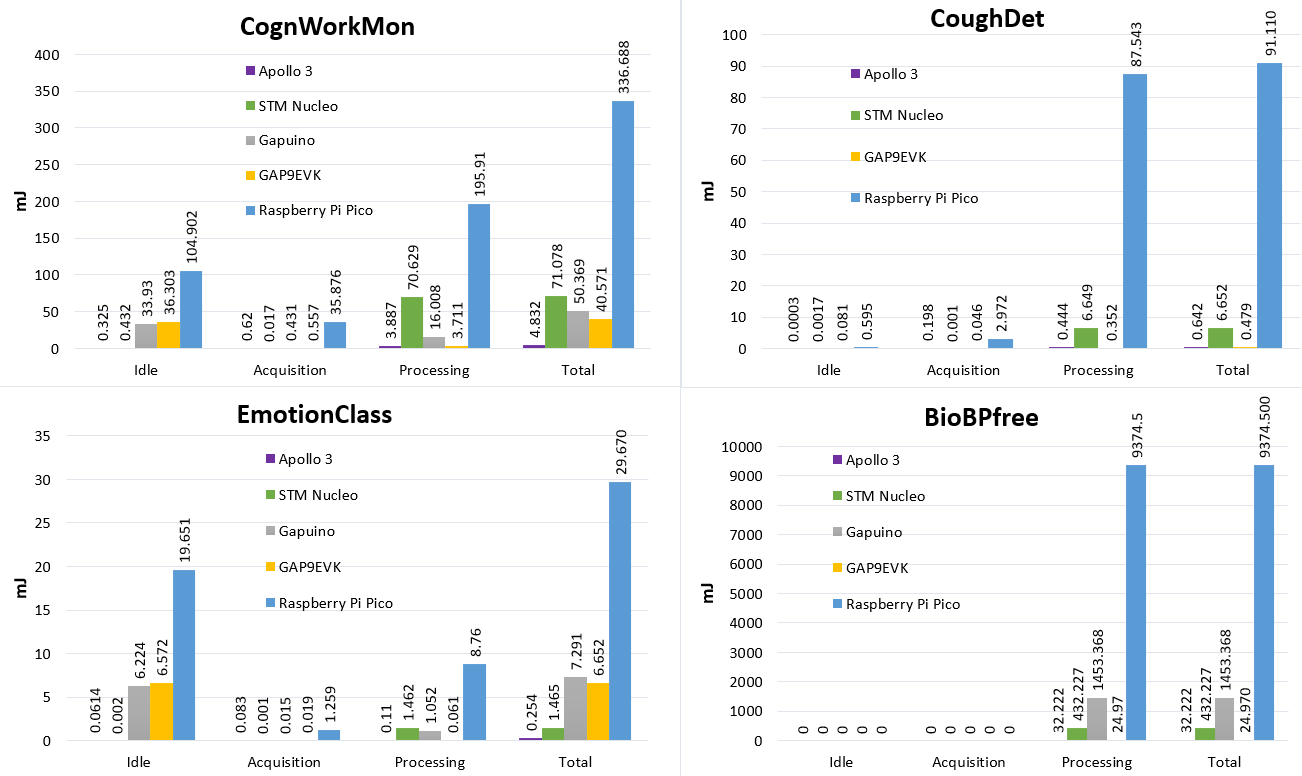}
		\caption{Single-core only applications}
	\end{subfigure}
    \caption{Energy (\si{\milli\joule}) per phase and platform.}
    \label{fig:energy}
 \end{figure}
%%%%%%%%%%%%%%

\subsection{Energy comparison}\label{sec:ExpRes:Energy}
In Fig.~\ref{fig:energy}, we present an energy graph for the seven BiomedBench applications ported to the selected low-power platforms. The graphs include the energy footprint of the five boards during the three typical phases and the total operating window. For applications with a multicore version, we report two instances for Gapuino, GAP9EVK, and Raspberry Pi Pico: one for single-core execution, referred to as -S, and one for the multicore execution, marked as -M. 

\subsubsection{Processing Phase}\label{sec:ExpRes:Energy:Proc}
In the processing phases, we observe that Ambiq Apollo3, Gapuino, and GAP9EVK have the most energy-efficient \glspl{mcu}. GAP9 is the most efficient for most applications thanks to the ability of the architecture to efficiently handle all types of workloads, as shown in Sect.~\ref{sec:ExpRes:Cycles}. Interestingly, depending on the application, we see variations in the relative processing energy between GAP8 and Apollo. GAP8 runs more efficiently \gls{seizdetcnn}, thanks to the \gls{dsp} extensions of its cores and the high degree of parallelism of the application. For the rest of the workloads, Apollo edges Gapuino, with the most significant difference in the case of \gls{gcl} as GAP8 lacks an \gls{fpu}. Raspberry Pi Pico exhibits an order of magnitude higher processing energy than Apollo~3 and Gapuino in all applications.

Surprisingly, the tests revealed a significant difference between STM32L4R5ZI and Apollo3 Blue \glspl{mcu}, although both share the same ARM Cortex-M4 processor architecture. Both platforms include \glspl{ldo} to reduce the voltage level. The power measurement test point in Nucleo-L4R5ZI is set at \SI{3.3}{\volt}, while Apollo bypasses the first \gls{ldo} on the board, providing a measurement point at \SI{1.8}{\volt}. In addition, Apollo incorporates a more efficient internal Buck converter. These distinctions could justify a difference of up to $3\times{}$ in energy consumption between both platforms. The difference in power consumption could also be explained by the \gls{ic} manufacturing process. Apollo uses TSMC \SI{40}{\nano\meter} low-power \cite{apollonm}, although we were unable to find the technology STM used in the \gls{ic} manufacturing to confirm our hypothesis.

It is worth highlighting the energy reduction of the multicore implementations in applications with a high degree of parallelism. In \gls{seizdetcnn} and \gls{gcl}, the multicore execution in GAP8 and GAP9 (8 and 9 cores, respectively) shows an energy reduction ranging from 48\% to 67\% compared to the single-core execution, respectively. In the case of Raspberry Pi Pico, which utilizes only two cores, energy reduction ranges from 24\% to 42\%. These results confirm the findings on multicore energy savings presented in \cite{Eli_multicore}. \textbf{\textit{Takeaway~8: Multicore execution yields significant energy savings in biomedical applications thanks to their inherent high degree of parallelism.}}

\subsubsection{Acquisition Phase}\label{sec:ExpRes:Energy:Acq}
STM has the most energy-efficient implementation in the acquisition phase, keeping the architecture in sleep mode while the \gls{dma} executes the data transfer. Gapuino is the second most efficient, also using \gls{dma}. Apollo~3 is more efficient than GAP9 despite not using \gls{dma}. GAP9 has a less optimized sleep mode while performing \gls{dma} transfers from peripherals to memory. Finally, Raspberry Pi Pico has at least one order of magnitude higher energy consumption than GAP9. Overall, the signal acquisition phase is not critical for energy efficiency, as it represents less than \SI{1.5}{\percent} of the total energy footprint in three of the five applications. In \gls{cwm}, acquisition consumes a maximum of \SI{13}{\percent} of the total energy in the case of Apollo~3. In \gls{gcl}, acquisition consumes up to \SI{25}{\percent} in the case of GAP9 due to inefficient sleep mode during acquisition, fast computations, and a high sample rate. \textbf{\textit{Takeaway~9: Applications with high acquisition intensity necessitate the presence of \gls{dma} and an efficient sleep mode during its operation in signal acquisition.} }

\subsubsection{Idle Phase}\label{sec:ExpRes:Energy:Idle}
The consumption in deep-sleep mode is critical to the energy footprint of each board. The lower the application duty cycle, the higher the impact of idle consumption. This observation is pronounced in the \gls{seizdetsvm} benchmark, where the sleeping phase accounts for up to \SI{97}{\percent} of the total energy consumption. However, the percentage of idle consumption gradually drops as we examine applications with higher duty cycles. In \gls{gcl}, sleep-mode consumption accounts for less than \SI{6}{\percent} on all boards. \textbf{\textit{Takeaway~10: Applications with low duty cycle necessitate an optimized deep-sleep mode.} }

\subsubsection{Overall Energy}
We summarize our observations on the total energy footprint in four key points. First, GAP9 is the most energy-efficient architecture during processing thanks to its \gls{dsp} extensions, hardware loops, and multicore acceleration, excelling in computationally heavy and loop-intense applications (i.e., \gls{seizdetcnn}, \gls{gcl}). However, it suffers from a poor retentive deep-sleep mode (i.e., \gls{seizdetsvm}, \gls{hcl}, \gls{ecl}). Second, Apollo~3 and STM32L4R5ZI have the best deep-sleep modes and perform the best in low-duty-cycle applications (i.e., \gls{seizdetsvm}, \gls{hcl}, \gls{ecl}). Third, Apollo~3 is the best all-around \gls{mcu}, also featuring a low energy footprint during processing and, hence, dominates its competitors in medium-duty-cycle applications (i.e., \gls{cwm}). Fourth, STM32L4R5ZI features the most energy-efficient acquisition mode (\gls{dma} and light sleep mode). Nevertheless, acquisition energy is not impactful in overall energy, as exemplified in the most acquisition-intensive applications (\gls{gcl}, \gls{coughdet}).

\subsection{Summary}\label{sec:ExpRes:Summary}
Multiple observations stem from analyzing the energy and performance results of running BiomedBench on different platforms. In this section, we emphasize the most important findings of our work.

\subsubsection{\textbf{Microarchitecture}}
To achieve high performance and, hence, improve the energy efficiency of platforms in the low-power wearables domain, hardware designers should consider the following microarchitectural features: 
\begin{itemize}
\item \textit{Barrel shifter}: essential for executing \gls{fxp} biomedical applications as shown in Section~\ref{sec:ExpRes:Instructions}.
\item \textit{Size of register file}: A register file of 16 registers can negatively impact performance in kernels such as \acrfull{mf}, as highlighted in \ref{sec:ExpRes:Perf:HCL}. However, for seven of the eight benchmark applications, the 32-register processors (CV32E40P GAP8 and CV32E40P GAP9) do not showcase any performance advantage over the 16-register processors (Arm Cortex-M0+, Arm Cortex-M4).
\item \textit{Support for 32-bit \gls{fxp} multiplications}: single-cycle execution of 32-bit integer multiplication with 64-bit result is vital for high performance in \gls{fxp} biomedical applications as shown in Section \ref{sec:ExpRes:Perf:SVM-CWM}.
\item \textit{\gls{dsp} microarchitectural extensions and hardware loops}: single-cycle \gls{mac} operations and zero-overhead hardware loops are strong assets for computationally hungry and loop-intensive applications as shown in Sections \ref{sec:ExpRes:Perf:CNN} and \ref{sec:ExpRes:Perf:ECL}. 
\end{itemize}

\subsubsection{\textbf{Low-power hardware features}}
To improve the energy efficiency of low-power wearables, hardware designers should focus their efforts on:
\begin{itemize}
\item \textit{Processing efficiency}: an optimized microarchitecture with \gls{dsp} \gls{isa} extensions suited to the characteristics of biomedical applications leads to faster and more energy-efficient processing, as shown in Section~\ref{sec:ExpRes:Energy:Proc}.
\item \textit{Efficient sleep mode that allows \gls{dma} operation}: can offer significant energy savings in applications with high acquisition intensity as shown in Section~\ref{sec:ExpRes:Energy:Acq}.
\item \textit{Optimized retentive deep-sleep mode}: dominant factor in applications with low duty cycle as shown in Section~\ref{sec:ExpRes:Energy:Idle}.
\end{itemize}

\subsubsection{\textbf{Software engineering}}
Software developers in the low-power wearables domain should pay attention to the following:
\begin{itemize}
\item \textit{Real-time interrupt-driven execution}: The application runtime and transitions between operating modes can be efficiently orchestrated through interrupts, as explained at the end of Section~\ref{sec:ExpSetup:Deployment}
\item \textit{Portability issues}: a real-time implementation of an end-to-end biomedical application is currently dependent on platforms' \gls{hal} as shown in Section~\ref{sec:ExpSetUp:portability}. To achieve portability in the future, FreeRTOS should be extended to integrate \gls{hal} functionality.
\item \textit{Transcendental computations}: can significantly hamper performance due to lack of advanced computational units in \gls{soa} low-power platforms as shown in Section~\ref{sec:ExpRes:Transcendental}.
\end{itemize}

\subsubsection{\textbf{\gls{soa} low-power platforms}}
Important observations on \gls{soa} platforms can hint at application developers searching for deployment platforms. Our most important takeaways here are:
\begin{itemize}
\item GAP9 dominates its competitors in raw execution speed thanks to its processing-efficient cores, increased operating frequency, and multicore cluster. It is the best choice for applications with tight real-time constraints, as shown in Section~\ref{sec:ExpRes:Time}.
\item GAP9 CV32E40P is the best microarchitecture since it can efficiently handle all computational workloads, as shown in Section~\ref{sec:ExpRes:Cycles}. The strongest features of this processor are the \gls{dsp} microarchitectural extensions (1-cycle 16-bit \gls{fxp} \gls{mac}, 1-cycle min/max set), the zero-overhead hardware loops, and the fast \gls{fpu}.
\item CV32E40P GAP9 dominates Arm Cortex-M4 in minimum/maximum search and sorting, featuring faster conditional assignment, specialized min/max instructions, and boosted by zero-overhead hardware loops, as shown in Sections~\ref{sec:ExpRes:Perf:HCL}, \ref{sec:ExpRes:Perf:ECL}. It also dominates Arm Cortex-M4 in loop-intensive 16-bit \gls{fxp} \gls{mac} operations, as illustrated in Sections~\ref{sec:ExpRes:Perf:CNN}.
\item Arm Cortex-M4 is the second best microarchitecture, featuring \gls{dsp} microarchitectural extensions but without zero overhead hardware loops and minimum/maximum find in one cycle. Arm Cortex-M4 may also suffer from the reduced 16-register register file in kernels like \acrfull{mf}, as shown in Section~\ref{sec:ExpRes:Perf:HCL}.
\item Arm Cortex-M0+ is significantly slower than the other processors lacking \gls{dsp} extensions and features a weaker instruction set compared to Arm Cortex-M4, as shown in Sec~\ref{sec:ExpRes:Cycles}. In this sense, Arm Cortex-M0+ is unsuitable for tackling the computational kernels of \gls{soa} biomedical applications.
\item GAP9, boosted by a multicore accelerator, is the most energy-efficient \gls{mcu} during processing and surpasses its predecessor Gapuino (Section~\ref{sec:ExpRes:Energy:Proc}). On the contrary, GAP9 suffers from poorly optimized sleep modes during idle and \gls{dma} acquisition (Sections~\ref{sec:ExpRes:Energy:Acq}, \ref{sec:ExpRes:Energy:Proc}), hence is suboptimal for extended idle periods. 
\item Apollo3 Blue is the most energy-efficient all-around \gls{mcu} featuring low energy footprint in active and idle phases, as shown in Section~\ref{sec:ExpRes:Energy}. 
\item STM32L4R5ZI has a very energy-efficient deep-sleep mode but a high current consumption during processing, as illustrated in Sections~\ref{sec:ExpRes:Energy:Proc},~\ref{sec:ExpRes:Energy:Idle}. It also has the most energy-efficient acquisition mode featuring an optimized sleep mode that allows the \gls{dma} to operate as shown in Section~\ref{sec:ExpRes:Energy:Acq}. 
\end{itemize}

%% file: Conclusions.tex
\section{Conclusions}\label{sec:conclusions}

This paper has presented a comprehensive analysis of the capabilities of state-of-the-art microcontrollers in the context of commercial wearable applications. In particular, this work has evaluated their processing capabilities, energy efficiency during idle, acquisition, and processing, and suitability for real-time requirements. Our findings shed light on the essential hardware design factors that directly influence performance in contemporary low-power designs. Additionally, we have provided actionable guidelines for software developers working with biomedical kernels, deploying applications on low-power \glspl{mcu}, and selecting optimal platforms for specific use cases. These insights not only underscored current best practices but also pointed out the way toward future advancements in biomedical wearables. More specifically, our analysis highlighted the need for continued innovation in energy-efficient processing architectures, which must be guided by application characteristics.